\documentclass[12pt]{aastex}

\def \h2         {\hbox{H$_2$}}

\def\approxlt{\lower.2em\hbox{$\buildrel < \over \sim$}}
\def\approxgt{\lower.2em\hbox{$\buildrel > \over \sim$}}

\begin{document}

\title {Submillimeter Follow-up of WISE-Selected
  Hyperluminous Galaxies}

\author {Jingwen Wu\altaffilmark{1},  
Chao-Wei Tsai\altaffilmark{2}, 
Jack Sayers\altaffilmark{3},
Dominic Benford\altaffilmark{4},
Carrie Bridge\altaffilmark{3},
Andrew Blain\altaffilmark{5},
Peter R. M. Eisenhardt\altaffilmark{1},
Daniel Stern\altaffilmark{1},
Sara Petty\altaffilmark{6},
Roberto Assef\altaffilmark{1},
Shane Bussmann\altaffilmark{7},
Julia M. Comerford \altaffilmark{8},
Roc Cutri\altaffilmark{2},
Neal J. Evans II\altaffilmark{8},
Roger Griffith\altaffilmark{2},
Thomas Jarrett\altaffilmark{2},
Sean Lake\altaffilmark{6},
Carol Lonsdale\altaffilmark{9},
Jeonghee Rho\altaffilmark{10},
S. Adam Stanford\altaffilmark{11},
Benjamin Weiner \altaffilmark{12},
Edward L. Wright\altaffilmark{6}, 
Lin Yan\altaffilmark{2}
}

\altaffiltext{1}{Jet Propulsion Laboratory, California Institute of Technology, 4800 Oak Grove Dr., Pasadena, CA 91109, USA; jingwen.wu@jpl.nasa.gov}
\altaffiltext{2}{Infrared Processing and Analysis Center, California Institute of Technology, Pasadena, CA 91125, USA}
\altaffiltext{3}{Division of Physics, Math and Astronomy, California Institute of Technology, Pasadena, CA 91125, USA}
\altaffiltext{4}{NASA Goddard Space Flight Center, Greenbelt, MD 20771, USA}
\altaffiltext{5}{Department of Physics and Astronomy, University of Leicester, LE1 7RH Leicester, UK}
\altaffiltext{6}{Department of Physics and Astronomy, University of California Los Angeles, Los Angeles, CA 90095, USA}
\altaffiltext{7}{Harvard-Smithsonian Center for Astrophysics, 60 Garden St. MS78, Cambridge, MA, 02138, USA}
\altaffiltext{8}{Department of Astronomy, University of Texas, Austin TX 78731,USA}
\altaffiltext{9}{National Radio Astronomy Observatory, 520 Edgemont Road, Charlottesville, VA 22903, USA}
\altaffiltext{10}{SETI Institute, 189 BERNARDO AVE, Mountain View, CA 94043 and SOFIA Science Center, NASA Ames Research Center, Moffett Field, CA 94035, USA}
\altaffiltext{11}{Department of Physics, University of California Davis, One Shields Ave., Davis, CA 95616, USA}
\altaffiltext{12}{Steward Observatory, 933 N. Cherry St., University of Arizona, Tucson, AZ 85721, USA}

\begin{abstract}

We have used the Caltech Submillimeter Observatory (CSO) to follow-up a sample of WISE-selected, hyperluminous galaxies, so called W1W2-dropout galaxies. This is a rare ($\sim 1000$ all-sky) population of galaxies at high redshift (peaks at $z$=2-3), that are faint or undetected by WISE at 3.4 and 4.6 $\mu$m, yet are clearly detected at 12 and 22 $\mu$m. The optical spectra of most of these galaxies show significant AGN activity.  
We observed 14 high-redshift ($z > 1.7$) W1W2-dropout galaxies with SHARC-II at 350 to 850 $\mu$m, with 9 detections; and observed 18 with Bolocam at 1.1 mm, with five detections. Warm {\it Spitzer} follow-up of 25 targets at 3.6 and 4.5 $\mu m$, as well as optical spectra of 12 targets are also presented in the paper. Combining WISE data with observations from warm {\it Spitzer} and CSO, we constructed their mid-IR to millimeter spectral energy distributions (SEDs). These SEDs have a consistent shape, showing significantly higher mid-IR to submm ratios than other galaxy templates, suggesting a hotter dust temperature. We estimate their dust temperatures to be $60-120$ K using a single-temperature model. Their infrared luminosities are well over 10$^{13}$\,L$_\odot$. These SEDs are not well fitted with existing galaxy templates, suggesting they are a new population with very high luminosity and hot dust. They are likely among the most luminous galaxies in the Universe. We argue that they are extreme cases of luminous, hot dust-obscured galaxies (DOGs), possibly representing a short evolutionary phase during galaxy merging and evolution. A better understanding of their long-wavelength properties needs ALMA as well as {\it Herschel} data.

\end{abstract}


\section{Introduction} \label{intro}

The redshift $z\sim$2-3 epoch stands out as a unique era for studying galaxy formation and evolution. At this epoch, the cosmic star formation rate reaches its peak (Heavens et al. 2004; Hopkins \& Beacom 2006;  Reddy et al. 2008), and ultra-luminous infrared galaxies (ULIRGs, $L_{IR} > 10^{12} L_{\odot}$, Sanders \& Mirabel 1996) contribute a significant fraction to the infrared luminosity density (Elbaz et al. 2002,  Chapman et al. 2005, Caputi et al. 2007, Reddy et al. 2008, Magnelli et al. 2009). The cosmic quasar density also peaks around $z\sim$2 (Hopkins et al. 2007, Assef et al. 2011). A framework of galaxy evolution through major mergers has been gradually built up by theorists (Barnes \& Hernquist 1992, Schweizer 1998, Jogee 2006, Hopkins et al. 2006, 2008). In one of the most popular scenarios (e.g. Hopkins et al. 2008), the tidal torques generated by major mergers funnel gas into the center of galaxies, leading to a central starburst and rapid growth of a supermassive black hole (SMBH). Black hole and supernova feedback terminate further star formation, evacuating the residual gas and dust, leaving a visible quasar and remnant spheroid. This picture establishes the evolutionary connections between ULIRGs, quasars, and massive elliptical galaxies.

Submillimeter galaxies (SMGs) are thought to be the analogues of local ULIRGs at high redshift (Blain et al. 2002, Tacconi et al. 2008). SMGs are selected by their strong cold dust emission at 850 $\mu$m ($F_{850} > $5 ~mJy). They are characterized by very high star formation rates (100-1000 M$_{\odot}$yr$^{-1}$) and infrared luminosity ($L_{IR}\sim 8\times 10^{12} L_{\odot}$, Chapman et al. 2005, Magnelli et al 2012). Although most SMGs host growing black holes (e.g., Alexander et al. 2005, 2008), their luminosities are normally dominated by star formation (Swinbank et al. 2004; Men\'{e}ndez-Delmestre et al. 2007, Younger et al. 2008, Hainline et al. 2011). The redshift distribution of SMGs strongly peaks at $z$=2-3 (Chapman et al. 2005), and the surface density of SMGs is several hundred per square degree. 

An 850$\mu$m selected sample (SMGs) may be biased toward ULIRGs with large amounts of dust, but miss a substantial population of ULIRGs with a smaller amount of (but warmer) dust, which can be found by surveys at shorter wavelengths. A series of surveys using bright {\it Spitzer} 24$\mu$m emission combined with optically faint photometry have been carried out to probe the ULIRG population with emission from smaller and warmer dust grains (e.g., Rigby et al. 2004, Donley et al. 2007, Yan et al. 2007, Farrah et al. 2008, Soifer et al. 2008, Lonsdale et al. 2009, Huang et al. 2009). One of the simplest search criteria is given as $F_{24} > 0.3$ mJy, and $R-[24]> 14$ (where $R$ and [24] are Vega magnitudes for $R$ band and {\it Spitzer} 24 $\mu$m), or roughly $F_{24}/F_{R} > 1000$ (Dey et al. 2008, Fiore et al. 2008), leading to a well defined $z \sim$2 population which is referred to as Dust Obscured Galaxies (Dey et al. 2008, hereafter D08). 
The most luminous DOGs have star formation rates (500-1000 M$_{\odot}$yr$^{-1}$ or more) and infrared luminosities ($L_{IR}\sim 10^{13} L_{\odot}$, Bussmann et al 2009, Tyler et al. 2009, Melbourne et al. 2012) that are comparable to SMGs. It has been proposed that both SMGs and DOGs are an early phase of galaxy merging, with SMGs representing an earlier, starburst-dominant phase, while luminous DOGs are in a transitional phase from starburst-dominated to AGN-dominated (e.g. Narayanan 2010). The bolometric luminosities also reach their maximum during these phases, making the most luminous galaxies in these phases also among the most luminous objects in the Universe. 

Looking for the most luminous galaxies in the Universe is one of the major goals of NASA's Wide-field Infrared Survey Explorer (WISE, Wright et al. 2010). WISE surveyed the entire sky at 3.4, 4.6, 12 and 22 $\mu$m (hereafter W1, W2, W3, W4) in 2010. The WISE dataset is well suited to investigate the starburst-AGN phase of distant, infrared luminous galaxies. At $z \sim$ 2-3, starburst- or AGN-heated hot dust can be traced by 12 and 22 $\mu$m emission, while the rest near infrared (NIR) obscuration is sampled by 3.4 and 4.6 $\mu$m continuum. Studies of luminous infrared galaxies with the WISE W1, W2 and W4 bands can take advantage of existing knowledge and techniques developed by earlier studies with {\it Spitzer} at similar wavelengths (IRAC at 3.6 and 4.5 $\mu$m, and MIPS at 24 $\mu$m). Observing W4 selected galaxies with WISE is similar to observing 24 $\mu$m bright galaxies with {\it Spitzer}, but with the surveyed area enlarged from a few tens of square degrees covered by existing DOG surveys to the entire sky. 

In order to search for hyperluminous infrared galaxies (HyLIRGs, $L_{IR} > 10^{13} L_{\odot}$) from the WISE dataset, the WISE team has explored multiple methods to select candidates. The most productive method so far has been to search for more heavily obscured galaxies, whose W1 (3.4 $\mu$m) and W2 (4.6 $\mu$m) flux densities become faint or undetected by WISE, while remaining easily detectable at 12 and/or 22 $\mu$m, with typical W4 (22 $\mu$m) flux densities $> 7 {\rm mJy}$. We call this population "W1W2-dropouts" (Eisenhardt et al 2012) or for brevity "W12drops".  Follow-up spectroscopy of more than 100 W12drop galaxies at large telescopes (this paper, Eisenhardt et al. (2012), see also Bridge et al. 2012) reveals that a large fraction ($> 65\%$) of these galaxies are at high redshift ($z > 1.5$), with the highest at $z$=4.6. Most of the redshifts are between 2 and 3, suggesting they also trace the peak epoch of cosmic star formation and QSO activity.  At these redshifts, such high flux densities at 22 $\mu$m imply extremely high luminosities. They are potentially hyperluminous galaxies. In order to understand the dust properties and calculate the total luminosities of these unusual galaxies, continuum measurements at longer wavelengths are crucial. As the first high redshift examples were identified, we began follow-up 0.35-1.1 mm continuum observations using the Caltech Submillimeter Observatory (CSO), in order to construct their SEDs and to explore the nature of W12drop galaxies. 

In this paper, we report the initial results of this follow-up study. The WISE data are described in section 2.1, and the W12drop population followed up with the CSO and reported here is listed in Table~1.  Sections 2.2 and 2.3 describe the CSO data,
while section 2.4 describes the follow-up optical spectroscopy, which is summarized in Table 2. 
Section 2.5 describes Spitzer follow-up observations at 3.6 $\mu$m and 4.5 $\mu$m of the W12drops, which
were selected to be difficult to detect by WISE at W1 (3.4 $\mu$m) and W2 (4.6 $\mu$m), and the photometry
for the sources is presented in Table 3.  Section 3 presents luminosity and dust temperatures constraints
from the photometry, while section 4 compares W12drop properties to those of DOGs and SMGs, 
and section 5 summarizes the findings. 
Throughout this paper we assume a ΛCDM cosmology with  $H_0 = 71$ km s$^{-1}$ Mpc$^{-1}$, $\Omega_m =
0.27$, and $\Omega_{\Lambda} = 0.73$ .

\section{Observations} \label{obs}
\subsection{WISE}

WISE began surveying the sky a month after it launched on 2009 December 14, completing its first coverage of the sky six months later, and continued surveying until February 1, 2011. The WISE all-sky data release was made public on 2012 March 14, and its content and characteristics are documented in the Explanatory Supplement\footnote{http://wise2.ipac.caltech.edu/docs/release/allsky/expsup}. WISE has achieved much better sensitivity than previous all-sky survey missions (5$\sigma$ point source sensitivities are better than 0.07, 0.1, 0.9 and 5.4 mJy in W1, W2, W3 and W4 bands (Explanatory Supplement \footnote[1]{Cutri et al. 2012, http://wise2.ipac.caltech.edu/docs/release/allsky/expsup/sec6\_3a.html}), and identified hundreds of millions of sources.

The selection criteria for W12drops use WISE catalog photometry, 
which provides PSF-ftting (i.e. total) magnitudes and uncertainties in the Vega system.  
The criteria are W1 $>$ 17.4, and either: a) W4 $< $7.7 and W2$-$W4 $>$ 8.2; 
or b) W3 $<$ 10.6 and W2$-$W3 $>$ 5.3). Additional details are given in Eisenhardt et al. (2012). 
This selection yields only about 1000 targets over the full sky. 
W3 and W4 flux densities and uncertainties for the W12drops discussed in this paper are provided in Table 3, 
converted from catalog magnitudes using zero-points of 29.04 and 8.284 Jy for W3 and W4 respectively (Wright et al. 2010).  
No color corrections have been made to these zero-points, because the results presented here are not sensitive to such corrections.

\subsection{SHARC-II}
The Submillimeter High Angular Resolution Camera II (SHARC-II) installed at the 10.4 m CSO telescope (Dowell et al. 2002) is a background-limited 350 and 450 $\mu$m facility camera,  and it is also equipped with a filter that allows 850 $\mu$m continuum observations. It adopts a ``CCD-style'' bolometer array with 12 $\times$ 32 pixels, resulting in a 0.97$\arcmin\ \times 2.59\arcmin\ $ field of view. The FWHM beam sizes of SHARC-II at 350 $\mu$m, 450 $\mu$m, and 850 $\mu$m are $8\farcs5$, $10\farcs$, and $19\farcs8$, respectively. We used SHARC-II to follow-up 14 high redshift W12drop galaxies during runs in July and September of 2010, and in February and September of 2011 (See Table 1). Most of the targets were only observed at 350 $\mu$m, with a few also observed at 450 $\mu$m and 850 $\mu$m. Examples of SHARC-II images of detected W12drop galaxies are presented in Figure 1.
Since the 350 and 450 $\mu$m atmospheric transmission is very sensitive to the weather, we only observed targets under good weather conditions (i.e. when the opacity at 225 GHz  $\tau_{\rm 225 GHz} < 0.06$, which occurs $\sim$ 20\% of the time). The targets and observing information are listed in Table 1. The Dish Surface Optimization System (DSOS, Leong 2006) was used to correct the dish surface figure for imperfections and gravitational deformations as the dish moved in elevation during observations. 

We used the Comprehensive Reduction Utility for SHARC-II (CRUSH, Kov\'{a}cs 2006), version crush2.01-4, to reduce the SHARC-II data. Option ``-deep'' in CRUSH was applied to optimize the signal-to-noise ratio (S/N) for faint ($<$ 100 mJy) point sources. Planets Uranus and Neptune, when available, were used for flux calibration, focus correction and pointing correction, and secondary calibrators such as K3-50 and CRL618 were used when these planets were not available. In order to flux calibrate, we used the starlink’s ``astro'' package to calculate the flux density of the calibrator within the proper beam size for a SHARC-II band, and the ``show'' package in CRUSH to obtain the readout flux density of the observed calibrator. The calibration factors (peaks) were derived by dividing the real peak flux density of the calibrator by the readout peak flux density from CRUSH within one beam. The flux density of the target and the rms noise were then derived by applying this calibration factor to the readout of the peak position of the target and off-peak positions 1-2 beams from the peak, using the CRUSH ``show'' package in the same way as for the calibrator, and convolving with the same beam size. The statistics of the calibration factor over all our runs indicates a calibration uncertainty of 20\%. We used the sweep mode for the SHARC-II observations, in which the telescope moves in a Lissajous pattern that keeps the central regions of the maps fully sampled. The edges are much noisier than the central regions, and to compensate for this, we used ``imagetool'' in CRUSH to eliminate the regions of each map that had a total exposure time less than 25$\%$ of the maximum. 

Pointing was checked every hour with planets and secondary calibrators. The pointing drift was normally less than 3$\arcsec$ in both the azimuth and zenith directions between pointing checks, and a pointing correction has been applied during the data stacking. The uncertainty on the centroid position of an object will be the quadrature sum of the pointing uncertainty and the measurement uncertainty on the centroid, which approximately equals to the beam FWHM divided by S/N. Given our SHARC-II detections normally have S/N of 3-4, and beam size at 350$\mu m$ is $8\farcs5$, the measurement uncertainty on the centroid for our targets is about 2$\arcsec$-3 $\arcsec$. Therefore, the uncertainty of offset between the SHARC-II centroids and WISE positions should be less than 5$\arcsec$-6 $\arcsec$, which is consistent with our SHARC-II detections (Figure 1).

\subsection{BOLOCAM}

Bolocam is a large format camera at the CSO with 144 detectors, a
circular eight-arcmin-diameter field of view (FOV), an observing band
centered at 1.1 mm, and a point-spread function with a 30 arcsec
full-width at half-maximum (FWHM) (Haig et al. 2004). 
Eighteen W12drop galaxies were observed with Bolocam during runs in
June 2010, December 2010, Febuary 2011, and September 2011, with observing
information listed in Table 1. The observations were made by scanning
the CSO in a Lissajous pattern, with scanning parameters chosen to keep
the source within the FOV 100\% of the time (Sayers et al. 2011) while scanning at an average
speed of 2$\arcmin$/sec in June 2010, December 2010 and February 2011 and an
average speed of 4 $\arcmin$/sec in September 2011.
A pointing model, accurate to 5$\arcsec$, was created
from frequent observations of bright objects within $\simeq 10$ degrees
of the target galaxies following the methods described in Sayers et al.
(2009). 
The flux calibration, in nV/Jy, was
determined according to the procedure described in Laurent et al.
(2005), 
based on observations of Neptune and K3-50A in June 2010, IRC 10216 in
December 2010, G34.3 and NGC 2071IR in February 2011,
and Uranus in September 2011 (Sandell 1994;  Griffin and
Orton 1993). The Uranus calibration model of Griffin and Orton 1993 was
updated based on the 143~GHz Bolocam results described in Sayers et al.
2012. We estimate the flux calibration to be accurate to $\simeq 10\%$.

Our atmospheric noise subtraction procedure was based on the
algorithms described in Sayers et al. (2011), 
with the following modifications: 1) an adaptive principal component analysis (PCA) 
algorithm\footnote{
 Due to an unknown change in the electromagnetic environment of
 the CSO prior to the December 2010 observations, a large
 fraction of the Bolocam detectors suffered from noise in
 several narrow spectral bands at the high-frequency end
 of the signal bandwidth. In addition
 to subtracting noise from atmospheric fluctuations, the adaptive
 PCA algorithm was effective at subtracting this pickup
 noise. This problem was resolved in April 2011,
 and allowed the faster scan speeds used in September 2011 (although
 we still used the same adaptive PCA algorithm to remove atmospheric
 noise from the September 2011 data). } was used to subtract the correlated
atmospheric signal over the FOV (Laurent et al. 2005, Downes et al. 2012)
and 2) the data were then high-pass filtered at a
characteristic frequency of 400 mHz, which corresponds to an angular
scale of 5$\arcmin$ and 10$\arcmin$ for the data collected at 2$\arcmin$/sec and 4$\arcmin$/sec.
As described in detail in Sayers et al. (2011),
the atmospheric
noise subtraction also attenuated the astronomical signal.
To account for this signal attenuation, observations
of the flux calibration sources were processed
in an identical way prior to determining the flux calibration.
Although the adaptive PCA algorithm is non-linear,
we verified via simulation that the combination of adaptive PCA and a
400 mHz high-pass filter results in a constant fractional amount
of signal attenuation for point-like objects with flux densities $<100$ Jy,
which is well above the flux density of any of our flux calibration sources.

We estimated the noise in our images via
jackknife realizations of our data following the
methods described in Sayers et al. (2011). These jackknifes involve multiplying a randomly selected subset of
half of our data by -1. By adding these jackknifes together we
preserve the noise properties of the data while removing the astronomical
signal.
In addition to instrumental and atmospheric noise, some
of our images also contain a non-negligible amount of
noise due to fluctuations in the astronomical
(back)foregrounds.
Using the galaxy number counts model in Bethermin \& Dole (2011),
we estimate the confusion noise (quantified by the RMS noise
fluctuations on beam size scales) to
be 0.6 mJy, which is approximately the confusion
noise measured at the same wavelength/resolution
with AzTEC (Scott et al. 2010).
The total uncertainty on the flux density of a
galaxy is then given by the quadrature sum of
instrument/measurement noise and confusion noise.
The noise fluctuations in the map are
Gaussian within our ability to measure them.
For our non-detections we quote 95\% confidence level upper limits based
on the formalism given in Feldman and Cousins (1998), who provided a
rigorous method for quoting upper limits from a measurement with Gaussian
noise and a physical constraint that the true underlying value is
non-negative (which is the case for our measurements, since negative flux
densities are unphysical). Specifically, our upper limits are computed
from the values given in Table 10 of Feldman and Cousins (1998).

\subsection{Optical Spectroscopy}

We obtained optical spectroscopy of the WISE-selected sources over
the course of four observing runs between July 2010 and February
2011.  Optical spectroscopic results for 14 sources in Table 1 are reported 
in Eisenhardt et al. (2012) and Bridge et al. (in preparation), as noted in Table 1. Table~2 lists the primary observing parameters, including
telescope, instrument, observing date, and integration time, for the remaining 12 W12drop galaxies. 
All of the targets were observed with multiple exposures, which were
generally dithered along the slit to improve the reduction quality.
Table~2 also lists the resultant redshifts, most of which are based
on multiple features and are therefore considered secure.

Most targets were observed with the dual-beam Low Resolution Imaging
Spectrometer (LRIS; Oke et al. 1995) on the Keck~I telescope.  All
of the LRIS observations used the 1\farcs5 wide longslit, the 5600
\AA\ dichroic, and the 400 $\ell\, {\rm mm}^{-1}$ grating on the
red arm of the spectrograph (blazed at 8500 \AA; spectral resolving
power $R \equiv \lambda / \Delta \lambda \sim 700$ for objects
filling the slit).  The July 2010 observations used the 600 $\ell\,
{\rm mm}^{-1}$ grism on the blue arm of the spectrograph ($\lambda_{\rm
blaze} =  4000$~\AA; $R \sim 750$), while the 2011 observations
used the slightly lower resolution 400 $\ell\, {\rm mm}^{-1}$ blue
grism ($\lambda_{\rm blaze} =  3400$~\AA; $R \sim 600$).  Observations
were generally obtained at a position angle that placed a brighter
offset star on the slit.  Since LRIS has an atmospheric dispersion
corrector, there are no issues with lost light due to observing at
non-parallactic angles.

The final 2 sources, W0220+0137 and W0116$-$0505, were observed with the Blue Channel
Spectrograph (BCS) on the 6.5-m MMT telescope in non-photometric
conditions on UT 2010 December 4.  These observations used the 1\farcs5
wide longslit and the 500 $\ell\, {\rm mm}^{-1}$ grating ($\lambda_{\rm
blaze} =  5400$~\AA; $R \sim 950$),  and were obtained
at a position angle of 47.3 $\deg$.

We processed the data using standard procedures, 
including bias subtraction, gain correction, cosmic ray removal, sky subtraction and
stacking the 2-dimensional spectra (e.g., Stern et al. 2010). 
The spectra were extracted using a 1\farcs5 aperture and
wavelength calibrated using internal arc lamps.   As a final step in
the wavelength calibration, we shifted the wavelength solution based
on telluric emission and absorption lines, conservatively providing
wavelength solutions that are robust to better than 1 \AA.  We flux
calibrated the spectra using observations of standard stars from
Massey \& Gronwall (1990), generally observed on the same night as the
science observations.  For photometric nights, we estimate that the flux
calibration is accurate to 10\%. For non-photometric data, which
includes the two sources observed with MMT, the spectrophotometry is
less accurate. 
The final, reduced spectra are presented in Figure 2.  The sources
range in redshift from $z = 1.990$ to $z = 3.592$ and all but three of
the sources (W0338+1941, W0926+4232 and W1830+6504) are clearly AGNs as evidenced by strong,
high-ionization emission lines such as \ion{O}{6}, \ion{C}{4},
and/or \ion{C}{3}].  One of the three outliers, W0338+1941, has an unusually broad Ly$\alpha$ profile, 
indicating that it is also likely an AGN. Note the diversity of spectroscopic features.
Most of the source spectra are dominated by strong, narrow
Ly$\alpha$ emission.  Some of the sources are clearly type-2 AGNs,
with only narrow emission features visible (e.g., W1422+5613).  

We highlight two sources with unusual spectra.  W0542$-$2705
shows a large number of moderate-width ($\sim 2400\ {\rm km}\, {\rm
s}^{-1}$) emission features.  Of particular note is the strong
\ion{Al}{3}~$\lambda 1857$ emission, which is stronger than the
\ion{C}{3}]~$\lambda 1909$ emission (\ion{Al}{3}/\ion{C}{3}] $\sim
2$).  In contrast, the corresponding line ratio for the Vanden Berk et al. (2001)
SDSS quasar composite is \ion{Al}{3}/\ion{C}{3} $= 0.02$,
approximately two orders of magnitude weaker.  The other unusual
spectrum is W0926+4232 which does not show any emission features,
though multiple absorption features as well as a Lyman forest break
clearly indicate a redshift of $z = 2.498$, analogous to the the spectrum of 
W1814+3412 reported in Eisenhardt et al. (2012).

\subsection{Warm {\it {\it Spitzer}} Follow-up}
Warm {\it {\it Spitzer}} observations at 3.6 and 4.5 $\mu$m of all the galaxies except one (W0211$-$2422) in Table 1 were obtained 
under program 70162 between November 2010 and July 2011. The sources were observed
using five exposures of 30s in each IRAC band (IRAC has a 5$\arcmin\ $ field of view with $1\farcs2$ pixels), with the medium scale Reuleaux dither pattern. The {\it {\it Spitzer}} pipeline post-BCD processed images, which are resampled onto $0\farcs6$ pixels, were used for photometry.
All targets in this paper with warm {\it {\it Spitzer}} follow-up were detected in both IRAC bands. Flux densities at 3.6 and 4.5 $\mu$m were measured in  $4\farcs8$ diameter apertures and are listed in Table 3. The aperture corrections are 1.205 and 1.221 for 3.6 and 4.5 $\mu$m, respectively. We did not apply a color correction, since the results presented here are not sensitive to such corrections. 
For W0211$-$2422, which was not observed by {\it {\it Spitzer}}, we list its W1 and W2 flux densities in Table 3. 

\section{Results} \label{results}

Of the 14 high redshift W12drop galaxies observed with SHARC-II at 350 $\mu$m, nine were detected at $\sim$ 3 $\sigma$ or above. The relatively high detection rate of these W12drop galaxies at 350 $\mu$m implies they are a submillimeter bright population with high infrared luminosity. We also observed three of the 14 targets using the SHARC-II 450 $\mu$m band, with one detection and one marginal detection (2$\sigma$-3$\sigma$); and observed one (W0410$-$0913, which has the brightest 350$\mu$m flux density) in the 850 $\mu$m band, with a detection. The flux densities of the detections are presented in Table 3.

We used Bolocam to follow-up 18 W12drop galaxies at z$>$1.7 in the 1.1 mm band, including six of the galaxies that were observed with SHARC-II. We obtained five detections, and 13 useful upper limits.  Flux densities of the detected targets and 95$\%$ confidence (2 $\sigma$) upper limits for undetected targets are presented in Table 3.
We also tabulate in Table 3 {\it {\it Spitzer}} 3.6 and 4.5 $\mu$m, and WISE 12 and 22 $\mu$m measurements. W0149+2350 was observed with the Submillimeter Array (SMA) at 1.3 mm (Wu, Bussmann, et al. in prep). We list this SMA measurement in the 1.1 mm column in Table 3.

\subsection{SEDs}

Taking the 3.6 $\mu$m to 1.1 mm measurements from Table 3, we construct the mid-IR to millimeter SED for W12drop galaxies. Figure 3 shows SEDs for the 9 SHARC-II detected W12drop galaxies compared with galaxy models.  In the first panel we overlay a wide range of templates at the corresponding redshift for one W12drop galaxy, normalized to the same W4 flux densities. These templates include the starburst-dominated galaxy Arp220, the AGN-starburst blend Mrk231, type I (unobscured), type II (obscured) AGN models (QSO 1 and QSO 2) from Polletta et al. (2007), a torus model (Polletta et al. 2006), and simulation models of DOGs (Narayanan et al. 2010). In the remaining panels we overlay only the Arp220 and Mark231 templates.  

The most notable feature is the apparently flat SEDs extending from the mid-IR to the submillimeter in all of these W12drop galaxies.  At 3.6 and 4.5 $\mu$m, large visual extinction must be added to the comparison templates to match the data, suggesting they are highly obscured. If we normalize all SEDs at their 22 $\mu$m flux densities, then the submm emission of W12drop galaxies is much fainter than expected for any other population, indicating their mid-IR to submm flux ratio is unusually high. 
Starburst models miss these SEDs by a large margin. QSO models are better, and the closest match is from the AGN dust torus model, but the fit is still poor. To quantitatively show the high mid-IR to submm ratio, we compare $\nu$L$\nu$(24$\mu m$)/$\nu$L$\nu$(350$\mu m$) in W12drops to this ratio in SMGs and DOGs in Figure 4a.

Figure 5 shows the SEDs of all submm detected W12drops plotted at their rest frame wavelength in $\nu$L$_{\nu}$ units, normalized by their total luminosity (see the next section). This figure shows a fairly consistent SED for all the W12drop galaxies reported in this paper. This SED has a power-law in the mid-IR (1-5 $\mu$m), a mid-IR bump that dominates the total luminosity contribution, and becomes flat in the mid-IR to submm. The typical SED of W12drops is quite different from any existing galaxy templates, indicating they may be a new type of galaxy. Their SEDs apparently peak at significantly shorter wavelengths than other galaxy templates, indicating hotter average dust temperatures.

\subsection{Luminosities and Temperatures} 
In order to understand the nature of the W12drop galaxies, we need to estimate their luminosities and dust temperatures. 
The standard method to do this is to fit several black-body models with wavelength-dependent opacities to fit SED points along a large range of wavelengths, to constrain both temperatures and luminosities. We already know that W12drop galaxies have unusually high mid-IR to submm ratios, and that the major luminosity comes from 24$\mu m$ to 350$\mu m$ emission (see Figure 5), so this is clearly the key wavelength range to characterize. At redshift 2-3, the IRAC [3.6] and [4.5] bands (rest wavelength 0.8-1.5 $\mu$m) may be significantly affected by stellar light, and at longer wavelengths, we see indications that these W12drop galaxies may have components in addition to a hot-dust component (see section 4). 

In many cases infrared-luminous galaxies are dominated by one major dust component, and can be approximated by a single-temperature modified blackbody model. For example, a single dust temperature model provides a good description of the far-IR and submm/mm SEDs of SMGs (e.g. Magnelli et al. 2012), with typical dust temperatures of 30-40 K (Chapman et al. 2005, Kov\'{a}cs et al. 2006, Coppin et al. 2008, Wu et al. 2009, Magnelli et al. 2012). Most DOGs can also be described by a single-temperature model with dust temperatures of 20-40 K (Melbourne et al. 2012).  Because the peak of the W12drop SEDs is not well sampled in the data presented here, it is unclear whether their 24$\mu m$ to 350$\mu m$ emission can be well described by a single-temperature model, and we are obtaining {\it Herschel} data to better determine this. In this paper we use a single-temperature model to describe the bulk of the emission from W12drops.  

We apply a single temperature, modified black-body model combined with a power-law model, to fit the mid-IR to mm SEDs.  At lower frequencies we use $S_{\nu} \propto \nu^{\beta} \times B_{\nu}(T)$, where $B_{\nu}(T)$ is the Plank function and $\beta$ is the dust emissivity index with $\beta=1.5$, attached smoothly to which at higher frequencies is a power-law with $S_\nu \propto \nu^{-\alpha}$. The two portions are joined at the frequency when the modified black-body slope equals that of the power-law ($\alpha$). The $\alpha$ parameter and dust temperature $T$ are constrained by W3, W4 and 350$\mu$m data, as well as 450$\mu$m, 850$\mu$m and 1.1 mm data when available. These data do not absolutely determine the shape of the SED, but the combination of the $\alpha$ and $T$ parameters provide a reasonable measure of the peak frequency of the SED.  In Table 4, we give the derived dust temperatures and mid-IR power-law indexes from this model.  

The derived dust temperatures of W12drops range from 60 K to 123 K, with a median value of 94 K. We applied this same model to calculate dust temperatures for DOGs in Bo$\ddot{\rm o}$tes field, using similar data from Melbourne et al. (2012). We used their reported  IRAC 8$\mu m$ and MIPS 24$\mu m$ data in place of W3 and W4 to calculate $\alpha$, together with their SPIRE 350 and 500 $\mu m$ data to estimate the dust temperature. We obtained similar dust temperatures to the 20-40 K reported by Melbourne et al. (2012). This confirms that W12drop galaxies are much hotter than typical DOGs. 

Our single-temperature model also provides a luminosity when $\alpha$, $\beta$ and T are fixed. However this luminosity is sensitive to the data points close to the peak, which is not well sampled here. Therefore instead, we use a simple, but conservative method to estimate the total luminosity, which is to simply connect the data points of all the available SED points with power-laws, and integrate the total flux densities.  This method may miss the luminosity close to the peak of the SED, so it provides a lower bound to the true luminosity. We list the conservative total luminosity calculated between 2-1000 $\mu$m in Table 4. The derived total luminosities range from 1.7 $\times 10^{13} L_{\odot}$ (W0211$-$2242), to 1.8 $\times 10^{14} L_{\odot}$ (W0410$-$0913), confirming that these galaxies are very luminous, well above the $10^{13} L_{\odot}$ threshold for HyLIRGs.

\section{Discussion } \label{sourcedisc}

Our CSO follow-up observations of 26 W12drop galaxies show that their luminosities are very high, with some over $10^{14} L_{\odot}$, and a median and mean of 5.7 and 6.1 $\times 10^{13} L_{\odot}$, all using the conservative power-law method. This is roughly an order of magnitude brighter than the typical SMG (with median luminosity $L\sim 8\times 10^{12} L_{\odot}$,  Chapman et al. 2005, Kov\'{a}cs et al. 2006), or DOG (with median and mean luminosity $\sim 6 \times 10^{12}$  and $9 \times 10^{12} L_{\odot}$, Melbourne et al. 2012), and is comparable to the brightest known optically selected quasars (Schneider et al. 2005). Extremely luminous infrared galaxies are often found to be magnified by galaxy-galaxy lensing (Blain 1996, Eisenhardt et al. 1996, Solomon \& Vanden Bout 2005, Vieira et al. 2010, Negrello et al. 2010).  An immediate concern about the hyperluminous W12drop galaxies is whether they are lensed, too. However, high-resolution imaging follow-up of selected W12drops does not indicate gravitational lensing (Eisenhardt et al. 2012, Bridge et al. in prep.), so that the derived $\sim 10^{14} L_{\odot}$ luminosities are consistent with being intrinsic based on the current data. Additional high-resolution follow-up observations with Hubble Space Telescope are currently underway, and should reveal if these W12drop galaxies are not lensed. If the lack of lensing is confirmed, these galaxies are one of the most luminous populations in the Universe.

Their unusually high dust temperatures and extremely high luminosities make W12drop galaxies of great interest for studying galaxy formation and evolution. How do they become so luminous? Are they experiencing special evolutionary events? What is their relationship to other well-established galaxy populations, such as SMGs and DOGs? 

Classical SMGs are defined with strong 850$\mu$m emission ($>$ 5mJy) which normally indicates significant cold-dust content. Table 3 gives examples of some W12drop galaxies that meet this criterion. Hence some W12drop galaxies would be selected as SMGs. However, the relatively low detection rate with Bolocam at 1.1 mm implies that many W12drop galaxies are not as bright as SMGs at longer wavelengths. This is understandable given that W12drop galaxies are dominated by emission from hotter dust. DOGs normally have both AGN and starburst contributions, with warmer dust grains than SMGs. In Dey et al (2008), DOGs are defined as galaxies with $F_{24} > 0.3$ mJy, and $R-[24]> 14$ (in Vega magnitudes), where the $R$ photometry is centered at 6393 \AA. Since the W4 band at 22 $\mu$m is similar to the {\it Spitzer} 24 $\mu$m band, our W4 $>$ 7 mJy selection corresponds to much higher flux densities at 24 $\mu$m than normal DOGs. To make a comparison between W12drops and typical DOGs, we obtained $r$-band (centered at 6231 $\AA$) photometry from SDSS (DR8), as listed in table 3, and used the $r$-W4 color to approximate the $R$-[24] color. Taking the average power-law index $\alpha$ of 2.09 from Table 4 and extrapolating to $r$-band and 24 $\mu$m, the difference between the $r$-W4 and $R$-[24] color ranges from 0.2 to 0.24 mag as $R$-[24] changes from 14 to 17. All 18 targets in Table 3 that are covered by SDSS DR8 meet the $r$-W4 $>$ 14 DOG criterion, with $r$-W4 ranging from 14.4 to 16.1 for $r$-band detected sources, and $r$-W4 $>$ 15.3 for $r$-band undetected sources (using $r$=22.9 Vega mag as the SDSS detection limit). In figure 6 we compare the distribution of $R$ vs. $R$-[24]  for these high-redshift W12drops to DOGs in D08. Clearly all W12drop galaxies in Table 3 qualify as DOGs, with similar colors, but are much brighter at 24 $\mu$m. 

Although many and maybe most W12drop galaxies can be classified as DOGs, their properties are quite different from normal DOGs. Comparing to the DOGs reported in D08, W12drop galaxies have an order of magnitude higher luminosity, although their redshift distributions are similar (Eisenhardt et al. 2012). Bussmann et al. (2009) used SHARC-II at 350 $\mu$m to follow-up a subset of DOGs with the brightest 24 $\mu$m flux densities from D08 (Fig 6), obtaining infrared luminosities (8-1000$\mu$m, $\sim 10^{13} L_{\odot}$) and dust temperatures ($>$30-60 K), still significantly lower than for the W12drop galaxies reported here. Since the D08 survey covered only $\sim $ 9 deg$^{2}$ (Bo$\ddot{\rm o}$tes field), the DOG surface density is $\sim$ 320 DOGs per square degree with 24 $\mu$m fluxes density greater than 0.3 mJy. The W12drop selection requires W4 $>$ 7 mJy, which is at the high 24 $\mu$m flux density end of the D08 sample, and only selects $\sim$ 1000 targets over the whole sky. The typical DOGs are 20 times fainter than the W12drops, but the latter are about 10000 times rarer. The high-z W12drop galaxies are apparently extreme cases of DOGs with very high dust obscuration and hotter dust temperatures, and appear to be hyperluminous, hot DOGs.

The very low surface density of W12drop galaxies suggests either they are intrinsically extremely rare, or are only seen during a very short phase of galaxy evolution.
DOGs are thought to be the transitional phase of mergers between starburst-dominated mode and AGN-dominated modes. Based on the mid-IR SED (3.6-24 $\mu$m), D08 classified DOGs into two categories, those which have a distinct ``bump'' in their SED between 3-10 $\mu$m attributed to the redshifted starlight from rest-frame 1.6 $\mu$m, and those whose mid-IR SED is a power-law. Bump DOGs are thought to be dominated by starbursts (Yan et al. 2005, Sajina et al. 2007, Farrah et al. 2008, Desai et al. 2009), and tend to have fainter 24 $\mu$m flux densities (Dey et al. 2008), while power-law DOGs are thought to be dominated by AGN in the mid-IR (Weedman et al. 2006, Donley et al. 2007, Yan et al. 2007, Murphy et al. 2009), and make up most of the bright end of the 24$\mu$m flux density distribution.  The fraction of power-law DOGs increases from 10\% at $F_{24 \mu m}=0.3$ mJy to 60\% at $F_{24 \mu \rm m}= 1$ mJy in the {\it Spitzer} Deep, Wide-Field Survey (Ashby et al. 2009). The mid-IR (MIPS 24 $\mu m$) to submm (SPIRE 250 $\mu m$) flux density ratio for power-law DOGs is found to be similar to the AGN dominated ULIRG Mrk231 (Melbourne et al. 2012). The IRAC1 to W4 SEDs of W12drops (Fig. 5) are more like the mid-IR SEDs of DOGs rather than SMGs' (Hainline et al. 2009). They show typical power-law shapes with no obvious bumps, and are very bright at 24 $\mu$m. Consequently, it is plausible that W12drop galaxies are also dominated by very powerful AGNs. These powerful, highly obscured AGNs can heat the surrounding dust cocoon to a very high temperature. 

Although the SEDs of W12drops are dominated by emission from very hot dust components (Figure 5) that are likely contributed by powerful AGNs, a hot AGN component alone can't explain all of the observed SED from the mid-IR to millimeter bands. It is likely that the SEDs are composed of multiple components with different temperatures. A more detailed model to decompose SEDs with multiple-temperature components needs a more complete set of SED data, which will become available from our ongoing {\it Herschel} program. But the 350$\mu$m and 1.1 mm data reported in this paper can give a useful constraint on the coldest component, if we assume that the 350 $\mu$m to 1.1 mm SED is tracing the coldest dust in these galaxies. In figure 7, we plot the modeled flux density ratios of 350 $\mu$m to 1.1 mm continuum versus the redshifts, for models with a single-temperature black-body times a wavelength-dependent opacity, with various dust temperatures and emissivities. W12drop galaxies with available 350 $\mu$m and 1.1mm measurements are plotted in the figure. For comparison, models based on galaxies with significant starburst components (Arp220 and Mrk231) are also plotted. 
For $\beta$ between 1.5 and 2.0, the 350 $\mu$m to 1.1 mm ratios of W12drop galaxies in Figure 7 apparently favor a model with T$_{\rm dust}$ less than 50 K, in addition to the $\sim$ 100 K hot dust component that dominates the mid-IR.  This temperature of colder dust is comparable to the typical dust temperature of $\sim$35 K associated with starburst galaxies (such as Arp220). Considering that the very hot AGN component will contribute more to the continuum at 350$\mu$m than at 1.1mm, the actual 350 $\mu$m to 1.1 mm ratios that trace the coldest dust could be lower, and therefore closer to the track of Arp220 or Mrk231 in Figure 7. This may imply that the cold dust component in these galaxies is not very different from those in starburst galaxies, and is possibly related to star formation. For example, in the detailed study of the first discovered W12drop galaxy (W1814+3412, Eisenhardt et al. 2012), a significant starburst is found, although only contributing a small fraction to the overall luminosity. The cold dust properties of W12drop galaxies may also be different from known obscured QSOs (e.g. Mart\'{i}nez-Sansigre et al. 2009). The ratio of 350 $\mu$m to 1.1 mm in AMS16, a high-z obscured quasar, is lower than for w12drops, as plotted in Figure 7. A detailed study of the long-wavelength properties for these W12drop galaxies, for instance, to distinguish the contribution and distribution of cold dust (star formation) and hot dust (AGN), will need observations from ALMA as well as {\it Herschel}.

The similarity between the optical to 22 $\mu$m SEDs of DOGs and W12drop galaxies, with the latter being much brighter, suggests W12drops may be the high luminosity tail of the DOG distribution. But the high mid-IR to submm luminosity ratio of W12drops implies they are much hotter than typical DOGs. Are these W12drops merely luminous DOGs, or a distinct population? Or do they have any evolutionary connection? Some theoretical models for DOGs (e.g. Narayanan et al. 2010) propose that SMGs, bump DOGs, and power-law DOGs may form an evolutionary sequence, representing the transition of merging galaxies from a starburst-dominated phase to an AGN-dominated phase, although direct observational support for this is still rare. In Figure 4a, we see a strong correlation between the mid-IR flux density and the mid-IR to submm luminosity ratio that supports such a sequence, with W12drops at the highest luminosities. The correlation is roughly linear, suggesting the cold dust component (traced by 350$\mu m$ emission which may be from a starburst) doesn't change significantly during this process, as clearly shown in Figure 4b, while the hot dust component (traced by 24$\mu m$) becomes stronger, possibly tracing the growth of an embedded SMBH. In this scenario, W12drops represent a late phase of this evolution, with more massive SMBHs and similar cold-dust components to SMGs and DOGs. If so, the low surface density of W12drops suggests either such a phase is very short, or not every galaxy goes through this stage. A better understanding of whether the populations have an evolutionary connection will need a fuller study of the W12drop population, and of the luminosity function of all these populations.

\section{Summary} \label{summary}

WISE has discovered a possibly new type of object, the W12drop galaxies. The results of our CSO submm/mm follow-up observations for a subsample of W12drop galaxies are as follows:

1. We observed 14 $z > 1.7$ W12drop galaxies with SHARC-II at 350$\mu$m, and nine were detected. We also observed 18 with Bolocam at 1.1 mm, and five were detected. 

2. The SEDs constructed from WISE, warm {\it Spitzer} and CSO data reveal consistent features for W12drop galaxies. These SEDs show a power-law shape in the mid-IR, and are apparently flat from the mid-IR to submm/mm. Their SEDs have unusually high mid-IR to submm luminosity ratios, indicating a hotter dust temperature than other populations. Their SEDs can not be well fit with existing galaxy templates, indicating they are likely a new population.

3. Using power-laws to connect the SED data points, we estimate their total luminosities to be at least 1.7 $\times 10^{13} L_{\odot}$ to 1.8 $\times 10^{14} L_{\odot}$. Using a single-temperature modified blackbody model with $\beta$=1.5, we estimate their dust temperatures to be 60 K to 120 K, much hotter than other infrared luminous galaxies. Besides the hot dust component, they may also have colder dust components that are similar to starburst galaxies. 

4. W12drop galaxies in this paper would also be selected as DOGs, but are at least 10 times more luminous and 10000 times rarer. They may be the extreme cases of very luminous, hot DOGs, and may represent a short evolutionary phase during galaxy merging, following the phase of SMG, bump DOG and power-law DOG.

\acknowledgements
J. Wu and R. Assef were supported by an appointment to the NASA Postdoctoral Program at the Jet Propulsion Laboratory, administered by Oak Ridge Associated Universities through a contract with NASA. N. Evans acknowledges support from NSF Grant AST-1109116. This publication makes use of data products from the Wide-field Infrared Survey Explorer, which is a joint project of the University of California, Los Angeles, and the Jet Propulsion Laboratory/California Institute of Technology, funded by the National Aeronautics and Space Administration. This work is based on observations made with the Caltech Submillimeter Observatory, which is operated by the California Institute of Technology under funding from the National Science Foundation, contract AST 90-15755. This work uses data obtained from the {\it Spitzer} Space Telescope, which is operated by the Jet Propulsion Laboratory, California Institute of Technology under contract with NASA. Some of the data presented herein were obtained at the W.M. Keck Observatory, which is operated as a scientific partnership among Caltech, the University of California and NASA. The Keck Observatory was made possible by the generous financial support of the W.M. Keck Foundation. Some data reported here were obtained at the MMT Observatory, a joint facility of the University of Arizona and the Smithsonian Institution. This paper uses data from SDSS (DR 8). Funding for SDSS-III has been provided by the Alfred P. Sloan Foundation, the Participating Institutions, the National Science Foundation, and the U.S. Department of Energy Office of Science. The SDSS-III web site is http://www.sdss3.org/.

\begin{deluxetable}{llrlcccc}
\tabletypesize{\scriptsize}
\rotate
\tablenum{1}
\tablewidth{7in}
\tablecaption{Targets observed with SHARCII and Bolocam}
\tablehead{
\colhead{Source} & \colhead{R.A.} & \colhead{Dec.} & \colhead{Redshift} &
\colhead{Band} &\colhead{UT date}& \colhead{Integration} & \colhead{$\tau_{\rm 225 GHz}$} \\
\colhead{Name $^{a}$} & \colhead{(J2000)} & \colhead {(J2000)} & \colhead{($z$)} &
\colhead{} & \colhead{}& \colhead{(hr)}  & \colhead{} }

\startdata
W0026+2015 &   00:26:09.24 & +20:15:56.2 &  1.990  &    350 $\mu$m &   2011 Sep 9     &  1.0     &   0.045   \\
W0116$-$0505 &   01:16:01.41 & -05:05:04.1 &  3.173 &    350 $\mu$m &   2011 Sep 10    &  1.0     &   0.045   \\
           &               &             &        &   1100 $\mu$m &   2011 Sep 18    &  2.3     & 0.15 \nl
W0149+2350 &   01:49:46.16 & +23:50:14.6 &  3.228$^{b}$ &    350 $\mu$m &   2010 Jul 28,31 &  2.0     &    0.05     \\
           &               &             &        &    450 $\mu$m &   2010 Sep 12    &  0.7     &    0.04      \nl
W0211$-$2242 &   02:11:34.63 & -22:42:23.4 &  1.746$^{b}$ &    350 $\mu$m &   2011 Sep 10    &  1.0     &    0.05   \\
W0220+0137 &   02:20:52.12 & +01:37:11.6 &  3.122 &    350 $\mu$m &   2010 Sep 12, 2011 Sep 10 & 1.8  &    0.045   \\
           &               &             &        &   1100 $\mu$m &   2011 Sep 19,20 &  3.7     &  0.16,0.07    \nl       
W0243+4158 &   02:43:44.18 & +41:58:09.1 &  2.010 $^{b}$ &    350 $\mu$m &   2011 Sep 10    &  0.8     &    0.05   \\
W0248+2705 &   02:48:58.81 & +27:05:29.8 &  2.210 &    350 $\mu$m &   2010 Sep 13, 2011 Sep 9 &  2.8  &    0.045    \\
           &               &             &        &   1100 $\mu$m &   2011 Feb 19,20,22, Sep 18-20 &  7.4  &0.15,0.07   \nl  
W0338+1941 &   03:38:51.33 & +19:41:28.6 &  2.123 &    350 $\mu$m &   2011 Sep 10    &  0.8     &    0.05   \\                                                
W0410$-$0913 &   04:10:10.60 & -09:13:05.2 &  3.592 &    350 $\mu$m &   2011 Feb 21, 2011 Sep 9 &  2.3  &    0.95, 0.045     \\
           &               &             &        &    850 $\mu$m &   2011 Feb 16-18    &  4.3  &    0.12     \nl  
           &               &             &        &   1100 $\mu$m &   2010 Dec 12-14    &  7.3  & 0.14,0.1     \nl
W0422$-$1028 &   04:22:48.82 & -10:28:32.0 &  2.227$^{b}$  &   1100 $\mu$m &   2011 Sep 17       &  2.7  & 0.21 \\
W0542$-$2705 &   05:42:30.90 & -27:05:40.5 &  2.532 &    350 $\mu$m &   2010 Sep 13       &  0.4  & 0.05      \\  
W0757+5113 &   07:57:25.07 & +51:13:19.7 &  2.277 $^{b}$&   1100 $\mu$m &   2011 Feb 18,22    &  4.0  & 0.21       \\
W0851+3148 &   08:51:24.78 & +31:48:56.1 &  2.640 &   1100 $\mu$m &   2010 Dec 13,14    &  5.3  & 0.13,0.1   \\
W0856+0005 &   08:56:28.08 & +00:05:48.7 &  2.519 $^{b}$ &   1100 $\mu$m &   2011 Feb 20       &  2.7  & 0.20 \\
W0859+4823 &   08:59:29.94 & +48:23:02.3 &  3.245$^{b}$&   1100 $\mu$m &   2010 Dec 13-14,2011 Feb 16-18   &  6.7  & 0.11,0.13 \\      
W0926+4232 &   09:26:25.44 & +42:32:51.9 &  2.498 &   1100 $\mu$m &   2011 Feb 19      &  4.0   & 0.28        \\
W1146+4129 &   11:46:12.87 & +41:29:14.3 &  1.772 $^{b}$&   1100 $\mu$m &   2011 Feb 16      &  3.0   & 0.15       \\
W1316+3512 &   13:16:28.53 & +35:12:35.1 &  1.956 $^{b}$&   1100 $\mu$m &   2011 Feb 19       &  2.7  & 0.21 \\
W1409+1335 &   14:09:25.56 & +13:35:02.1 &  3.048 $^{b}$&   1100 $\mu$m &   2011 Feb 17,22   &  2.3   & 0.1, 0.2      \\
W1422+5613 &   14:22:28.86 & +56:13:55.6 &  2.524 &   1100 $\mu$m &   2011 Feb 18,20,21 &  9.0  & 0.2         \\
W1603+2745 &   16:03:57.39 & +27:45:53.3 &  2.633 $^{b}$&    350 $\mu$m &   2010 Sep 13     &  0.5     &    0.04   \\   
W1814+3412 &   18:14:17.30 & +34:12:25.0 &  2.452 $^{c}$ &    350 $\mu$m &   2010 Jul 13,23  &  2.7     &    0.06, 0.04    \\  
           &               &             &        &    450 $\mu$m &   2010 Sep 12,13  &  1.8     &    0.045    \nl
           &               &             &        &   1100 $\mu$m &   2010 Jun 17,18  &  6.0     &    0.1     \nl
W1830+6504 &   18:30:13.53 & +65:04:20.5 &  2.653 &   350 $\mu$m &   2011 Sep 9,10    &  2.7     &    0.05   \\    
W1835+4355 &   18:35:33.71 & +43:55:49.1 &  2.298 $^{b}$&    350 $\mu$m &   2010 Sep 12     &  1.0     &    0.04    \\
           &               &             &        &    450 $\mu$m &   2010 Sep 13     &  0.6     &    0.04    \nl
W2207+1939 &   22:07:43.84 & +19:39:40.3 &  2.022 $^{b}$&   1100 $\mu$m &   2010 Dec 14     &  1.8     & 0.1        \\
W2238+2653 &   22:38:10.20 & +26:53:19.8 &  2.405 &   1100 $\mu$m &   2011 Sep 20     &  1.7  & 0.075 \\
\enddata
\tablenotetext{a}{According to the WISE source naming convention ($http://wise2.ipac.caltech.edu/docs/release/allsky/expsup/sec1\_6a.html$), targets reported here have formal source designations of: WISE Jhhmmss.ss$\pm$ddmmss.s. For example, the first target is WISE J002609.24+201556.2. But we use a brief form of names: Whhmm$\pm$ddmm, in this table and throughout the paper.}
\tablenotetext{b}{Spectral information will be reported in Bridge et al. in prep}
\tablenotetext{c}{Spectral information is reported in Eisenhardt et al. 2012}
\end{deluxetable}

\begin{deluxetable}{lllccl}
\tabletypesize{\scriptsize}
\tablenum{2}
\tablewidth{6.5in}
\tablecaption{Optical Spectroscopy}
\tablehead{

\colhead{Source} &

\colhead{Telescope/Instr.} &

\colhead{UT Date} &

\colhead{Exp. (s)} &

\colhead{$z$} &

\colhead{Notes}}

\startdata
W0026$+$2015  & Keck/LRIS         &  2010 Nov 08 & 600 + 300     &1.990 &  Ly$\alpha$, NV, CIV, HeII, [NeIV], MgII\\
W0116$-$0505   & MMT/BCS          &  2010 Dec 04 & $3 \times 600$ & 3.173  & Ly$\beta$,Ly$\alpha$,NV,SiIV/OIV \\
W0220$+$0137  & MMT/BCS          & 2010 Dec 04 & $3 \times 600$  & 3.122  & Ly$\beta$,Ly$\alpha$,NV,SiIV/OIV \\    
W0248$+$2705  & Keck/LRIS        & 2010 Nov 09 & $600 + 300$     & 2.210  & Ly$\alpha$,CII],MgII \\                
W0338$+$1941  & Keck/LRIS        & 2011 Feb 02 & $2 \times 900$  & 2.123  & Ly$\alpha$\\
W0410$-$0910  & Keck/LRIS        & 2010 Nov 08 & $2 \times 600$  & 3.592  & CIV,HeII (Ly$\alpha$ at dichroic) \\ 
W0542$-$2705  & Keck/LRIS        & 2010 Nov 08 & $600 + 300$     & 2.532  & CIII,OVI,Ly$\alpha$,NV,... \\          
W0851$+$3148  & Keck/LRIS        & 2011 Nov 08 & $600 + 300$     & 2.640  & OVI,Ly$\alpha$,CIV,HeII,CIIII],CII] \\ 
W0926$+$4232  & Keck/LRIS        & 2011 Feb 02 & $2 \times 900$  & 2.498  & Absorption lines only \\                                             
W1422$+$5613  & Keck/LRIS        & 2010 Jul 15 & $600 + 300$     & 2.524  & Ly$\alpha$,NV,CIV \\     
W1830$+$6504  & Keck/LRIS        & 2010 Jul 13 & $2 \times 600$  & 2.653  & Likely Ly$\alpha$\\          
W2238$+$2653  & Keck/LRIS        & 2010 Nov 08 & $2 \times 900$  & 2.405  & Ly$\alpha$, NV, SiIV/OIV , CIV\\

\enddata



\end{deluxetable}

\begin{deluxetable}{llcccccccc}
\tabletypesize{\scriptsize}
\tablenum{3}
\tablewidth{6in}
\tablecaption{Photometry of Targets$^{a}$}
\tablehead{
\colhead{Source}  &  \colhead{SDSS r $^{b}$} & \colhead{3.6$\mu$m }    & \colhead{4.5$\mu$m} &
 \colhead{12$\mu$m} &    \colhead{22$\mu$m} &
 \colhead{350$\mu$m} &   \colhead{450$\mu$m} &
 \colhead{850$\mu$m} &   \colhead{1100$\mu$m} 
\\
\colhead{} &  \colhead{(mag)} &\colhead{($\mu$Jy)} & \colhead{($\mu$Jy)} &
\colhead{(mJy)} & \colhead{(mJy)} & \colhead{({\rm mJy})} &
\colhead{(mJy)} & \colhead{(mJy)} & \colhead{{\rm (mJy)}} 
}
\startdata
W0026+2015 & 22.10 &40.1(2.0) & 80.5(2.0) & 2.77(0.13) & 15.05(0.97)  &  $<21$  &  ...      &   ...       & ...\\
W0116$-$0505 & 21.38 &50.7(2.2) & 89.4(2.3) & 2.39(0.13) & 12.96(1.00)  &  36(12) & ...         &  ...     &$<8.7$\\
W0149+2350 & $<$22.9 &19.7(1.7) & 34.7(1.5) & 1.77(0.10) & 9.18(0.76)   &  29(8)  & 35(9)    &  ...   &  2(0.4)$^{c}$\\
W0211$-$2242 & ‚Ä¶ &30.5(11)$^{d}$ & 78.0(10)$^{d}$ & 3.31(0.11) & 11.31(0.73)  &  56(15) &	...    &	...  & ... \\
W0220+0137 & 21.84 &25.2(1.8) & 38.4(1.4) & 1.78(0.10) & 11.98(0.81)  &  43(9)  &  ...        &  ...   &  6.2(2.0)\\
W0243+4158 & ... &23.1(1.7) & 70.7(2.0) & 2.56(0.13) & 9.02(0.93)   &  38(13) &  ...       &	 ...    & ... \\
W0248+2705 & ... &31.5(1.9) & 52.0(1.6) & 2.04(0.14) & 11.11(1.05)  &  32(8)  &  ...        & ...       &$<3.6$\\
W0338+1941 & ‚Ä¶ &12.2(1.6) & 37.5(1.6) & 1.97(0.14) & 10.27(0.99)  &  $<31$	 &   ...    &   ...     & ... \\
W0410$-$0913 & ... &26.7(1.8) & 46.1(1.5) & 2.45(0.14) & 12.35(0.99)  &  118(17)&  ...        &40(14)  & 13.6(2.6)\\
W0422$-$1028 & ... &18.0(1.6) & 55.0(1.6) & 2.74(0.13) & 10.70(1.02)  &   ...      &   ...       &  ...      &$<4.9$\\
W0542$-$2705 & ‚Ä¶ &24.6(1.7) & 29.2(1.2) & 2.55(0.11) & 14.08(0.90)  &   $<47$ &  ...    &	 ...    & ...\\
W0757+5113 & 22.27 &20.0(1.7) & 35.0(1.4) & 1.46(0.11) & 9.31(0.84)   &   ...      &   ...       &     ...   &$<4.7$ \\
W0851+3148 & 21.64 &41.4(2.0) & 88.5(2.2) & 3.51(0.16) & 14.73(1.02)  &   ...      &   ...       &   ...    & $<3.4$ \\
W0856+0005 & $<$22.9 &52.7(2.2) & 74.2(2.0) & 2.94(0.13) & 15.06(0.96)  &   ...      &   ...       &   ...    & $<9.4$ \\
W0859+4823 & $<$22.9 &16.4(1.6) & 44.8(1.4) & 2.22(0.10) & 11.83(0.91)  &   ...      &  ...        &   ...    & 6.2(1.5) \\
W0926+4232 & $<$22.9 &18.6(1.6) & 28.6(1.2) & 1.45(0.11) & 7.78(0.93)   &   ...      &  ...        &   ...     & $<5.5$ \\
W1146+4129 & $<$22.9 &23.6(1.7) & 45.7(1.5) & 3.90(0.13) & 20.35(1.05)  &   ...      &  ...        &   ...    &$<6.5$ \\
W1316+3512 & 22.56 &22.0(1.7) & 49.4(1.5) & 3.00(0.12) & 12.62(0.94)  &  ...       &  ...        &   ...    & $<14.2$ \\
W1409+1335 & $<$22.9 &6.9(0.4)  & 16.4(1.0) & 1.63(0.09) & 9.44(0.73)   &   ...      &  ...        &   ...     & 5.7(2.1) \\
W1422+5613 & $<$22.9 &28.6(1.8) & 74.3(1.9) & 3.07(0.09) & 11.95(0.68)  &   ...      &  ...        &   ...    & $<2.9$ \\
W1603+2745 & $<$22.9 &29.9(1.8) & 47.6(1.5) & 3.15(0.12) & 9.53(0.87)   &   $<56$  &   ...       &  ...     & $<27.9$\\
W1814+3412 & 23.00 &20.8(2.1) & 26.5(1.9) & 1.86(0.10) & 14.38(0.86)  &   33(9) &  $<32$     &   ...     &$<2.4$\\
W1830+6504 & $<$22.9 &14.2(1.6) & 41.9(1.4) & 2.25(0.05) & 7.56(0.34)   &   $<31$ & ...    & ...    &...\\
W1835+4355 & ... &51.5(2.2) & 142.8(3.0) & 6.13(0.13) & 27.05(0.87)  &   46(16)&  31(14)  & ...  & ...\\
W2207+1939 & $<$22.9 &41.7(2.0) & 57.8(1.7) & 1.49(0.11) & 10.27(0.90)  &   ...      &  ...        &   ...     & $<8.7$ \\
W2238+2653 & 22.77 &42.2(2.0) & 58.8(1.6) & 2.35(0.11) & 17.15(0.98)  &   ...      &  ...        &  ...      & 6.0(2.2) \\
\enddata
\tablenotetext{a}{Photometry of 3.6 and 4.5 $\mu m$ is from warm Spitzer; 12 and 22 $\mu m$ data is from WISE; 350, 450 and 850 $\mu m$ data is from CSO/SHARC-II, and 1.1 mm data  is from CSO/Bolocam. Numbers in parentheses are 1$\sigma$ uncertainties. For undetected targets, we give 95\% (2$\sigma$) upper limits at 350, 450, 850 and 1100 $\mu$m columns.}
\tablenotetext{b}{r band magnitude from SDSS (DR8). For undetected targets, we use 22.9 mag as upper limits.} 
\tablenotetext{c}{Flux density at 1.3mm, obtained from the SMA (Wu, Bussmann et al. in prep) }
\tablenotetext{d}{Data from WISE 3.4$\mu$m and 4.6$\mu$m measurements.}

\end{deluxetable}

\begin{deluxetable}{lrlc}
\tabletypesize{\scriptsize}
\tablenum{4}
\tablewidth{4.5in}
\tablecaption{Dust properties and luminosities for the sample galaxies with submm data}
\tablehead{

\colhead{Source} &

\colhead{T$_{\rm dust}$} &

\colhead{$\alpha$ $^{a}$} &

\colhead{Power-law Luminosity$^{b}$} \\

\colhead{} & 
\colhead{K} &
\colhead{} &

\colhead{L$_{\odot}$}
}

\startdata
W0116$-$0505 & 123$\pm$8 &  2.42$\pm$0.15  &    7.4 $\times 10^{13}$\\
W0149+2350 & 100$\pm$5 & 2.05$\pm$0.10  &   5.7 $\times 10^{13}$ \\
W0211$-$2242 &  60$\pm$5 & 1.73$\pm$0.10    &         1.7 $\times 10^{13}$ \\
W0220+0137 & 118$\pm$6 & 2.50$\pm$0.15 &  7.1 $\times 10^{13}$  \\
W0243+4158 & 68$\pm$5  &  1.81$\pm$0.20  &    2.0 $\times 10^{13}$ \\
W0248+2705 &  87$\pm$8 & 2.28$\pm$0.10	 &  2.8 $\times 10^{13}$ \\ 
W0410$-$0913 &  82$\pm$5 & 2.16$\pm$0.10 & 1.8 $\times 10^{14}$ \\
W1814+3412 & 113$\pm$7 & 2.89$\pm$0.10	 &  4.0 $\times 10^{13}$ \\      
W1835+4355 & 94$\pm$10 & 1.96$\pm$0.10	 &  6.5 $\times 10^{13}$ \\     

\enddata

\tablenotetext{a}{The power-law index assuming in the mid-IR $f_{\nu} \propto \nu^{-\alpha}$, for a single temperature, modified black-body model combined with a power-law model. See section 3.2.}
\tablenotetext{b}{Total luminosity ($\sim$2-1000 $\mu$m) calculated by connecting all available SED points with power-laws, which gives a lower bound of the total luminosity. See section 3.2.}

\end{deluxetable}

\clearpage

\begin{figure}
\epsscale{0.80}
\rotatebox{90}{\plotone{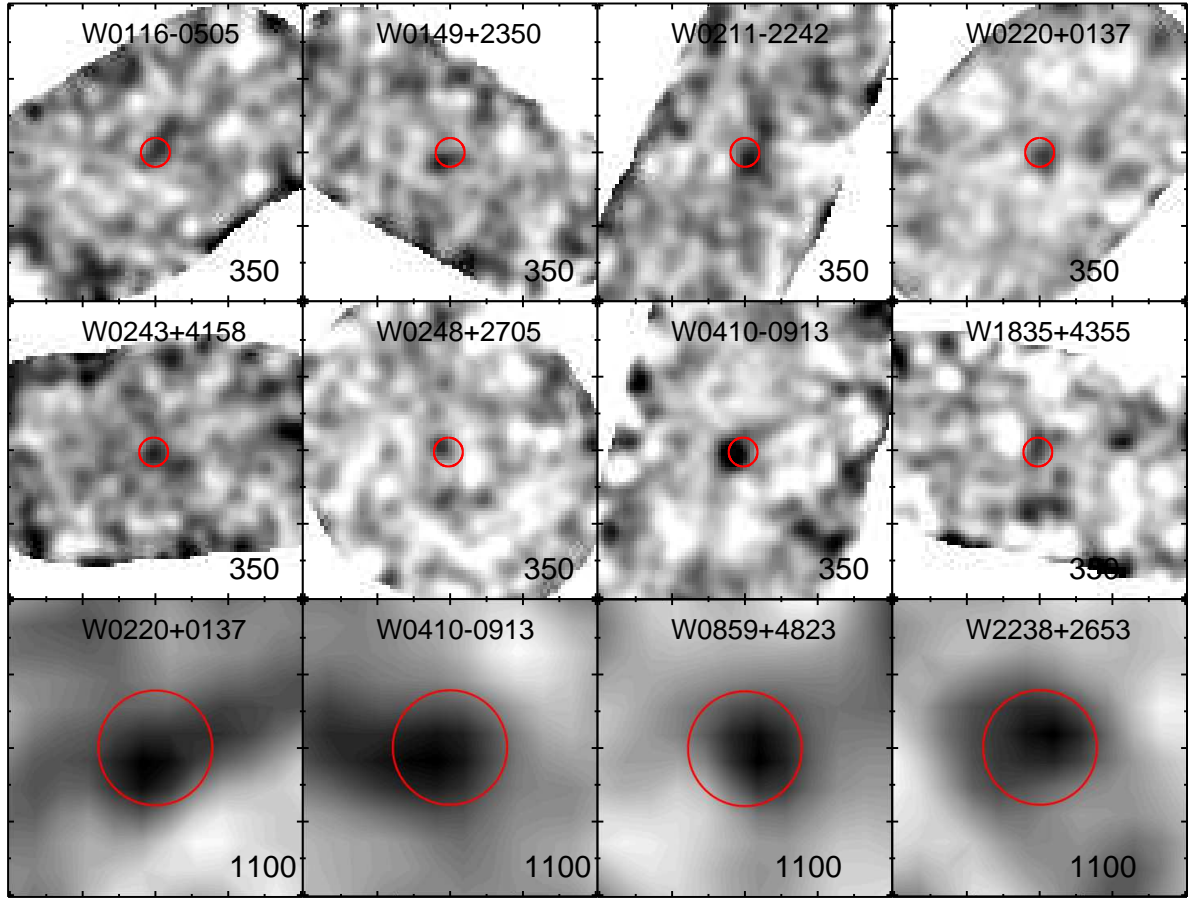}}
\caption{
Example SHARC-II and Bolocam images of detected W12drop galaxies. All panels are 2$\arcmin \times$ 2$\arcmin$.
The source name and observing band are listed in each panel. The center of the red circle marks the target position identified from WISE, and the sizes of the red circle represents the beam-smoothed resolution (FWHM) for each image, which is $11\farcs7$ for 350$\mu$m maps, and 42$\arcsec$ for 1.1mm maps.}
\end{figure}

\begin{figure}
\epsscale{0.70}
\rotatebox{270}{\plotone{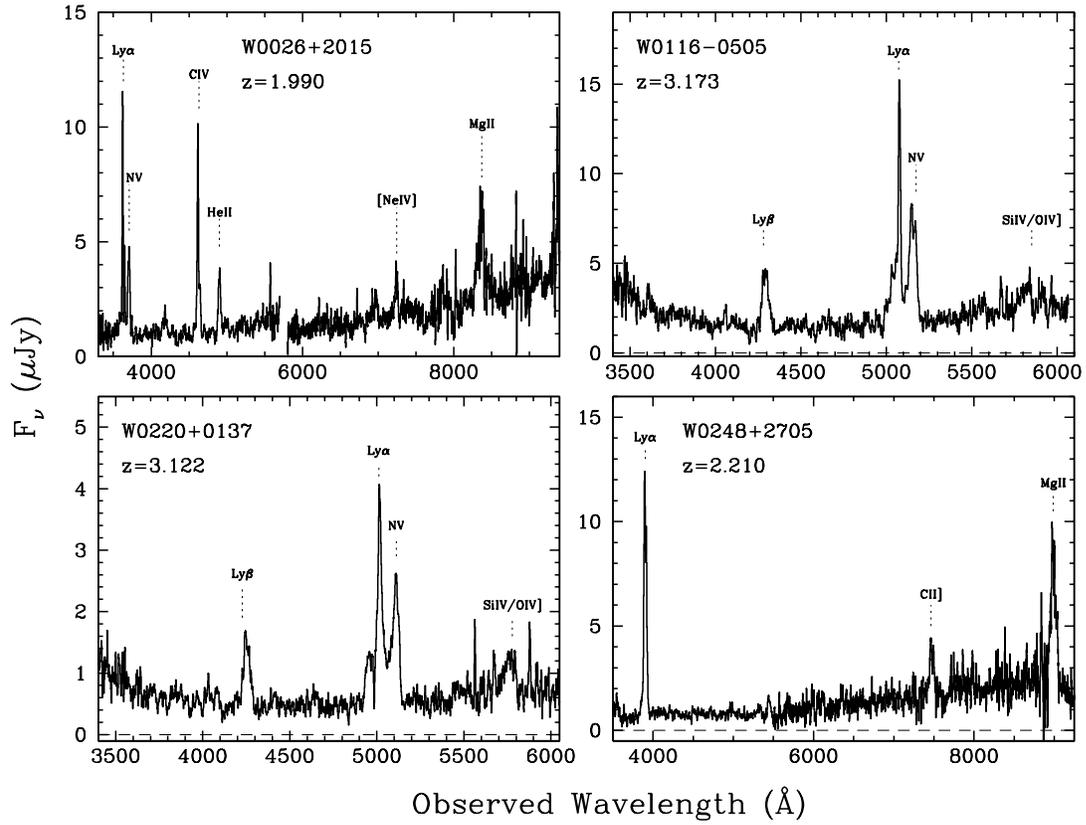}}
\caption{
Optical spectra of the 12 WISE-selected sources.  Prominent
spectroscopic features are labeled.
}
\figurenum{2}
\end{figure}

\begin{figure}
\epsscale{0.70}
\rotatebox{270}{\plotone{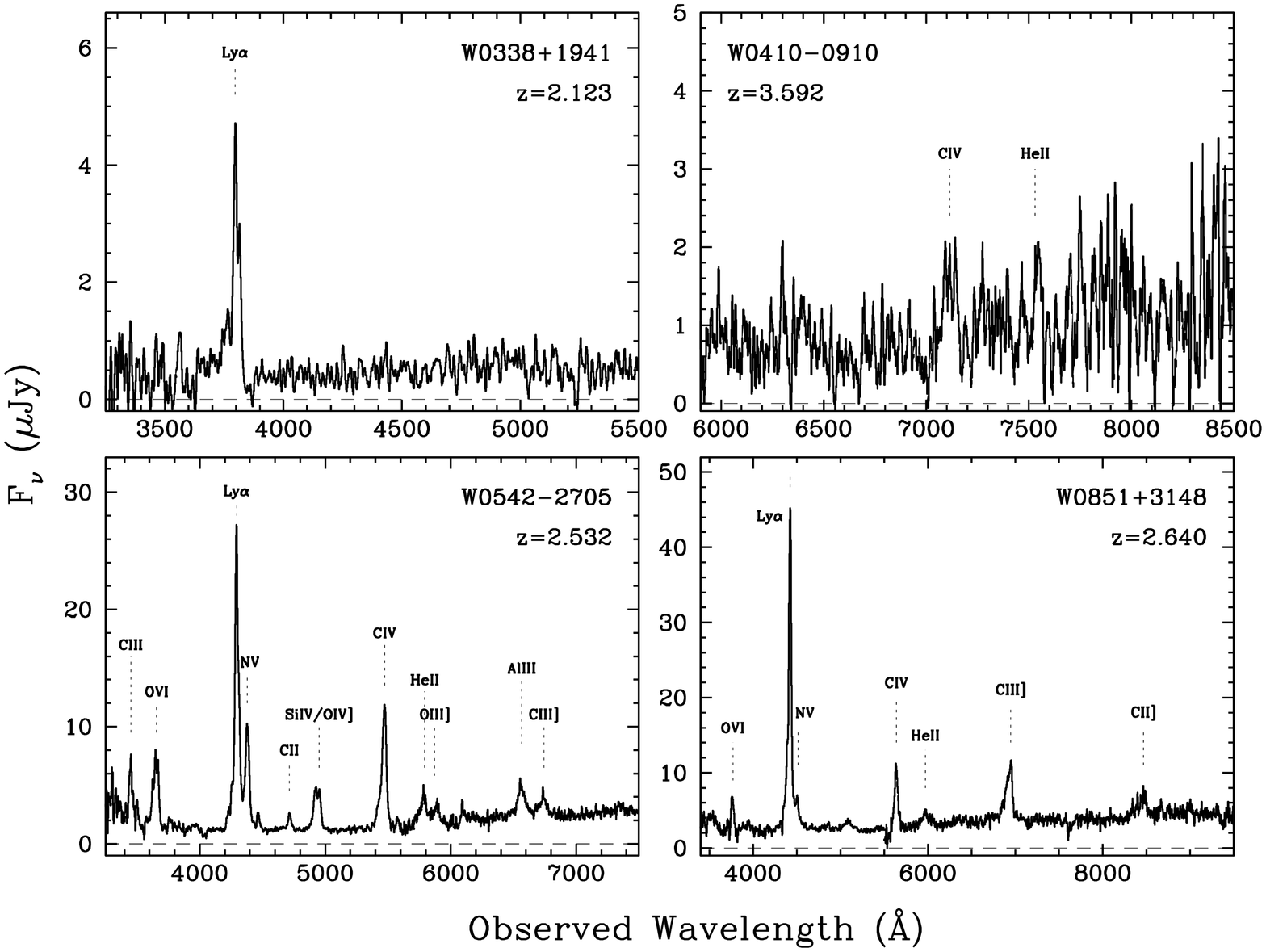}}
\caption{
Continued.
}
\figurenum{2}
\end{figure}

\begin{figure}
\epsscale{0.70}
\rotatebox{270}{\plotone{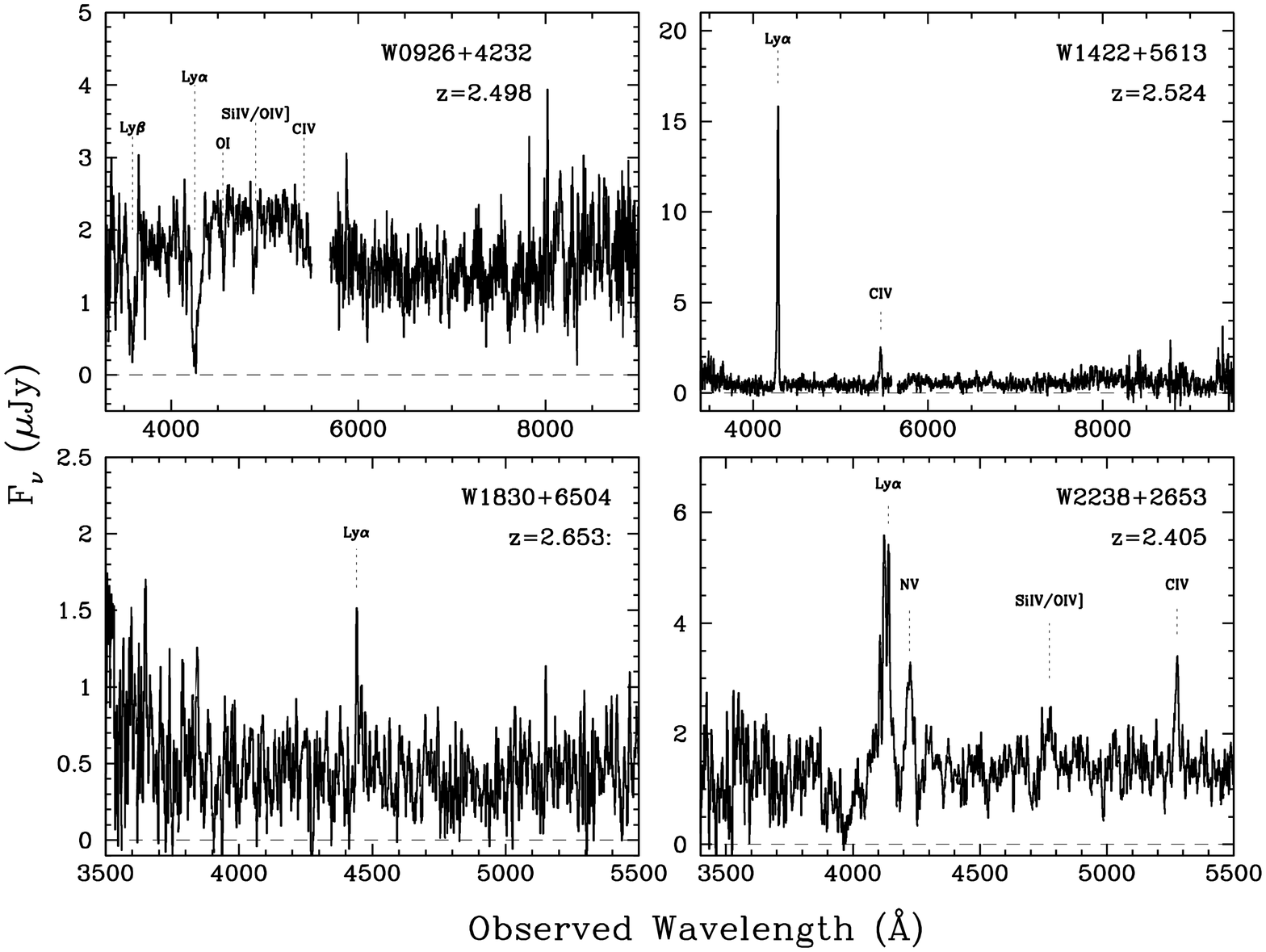}}
\caption{
Continued.
}
\end{figure}

\begin{figure}
\epsscale{0.9}
\plotone{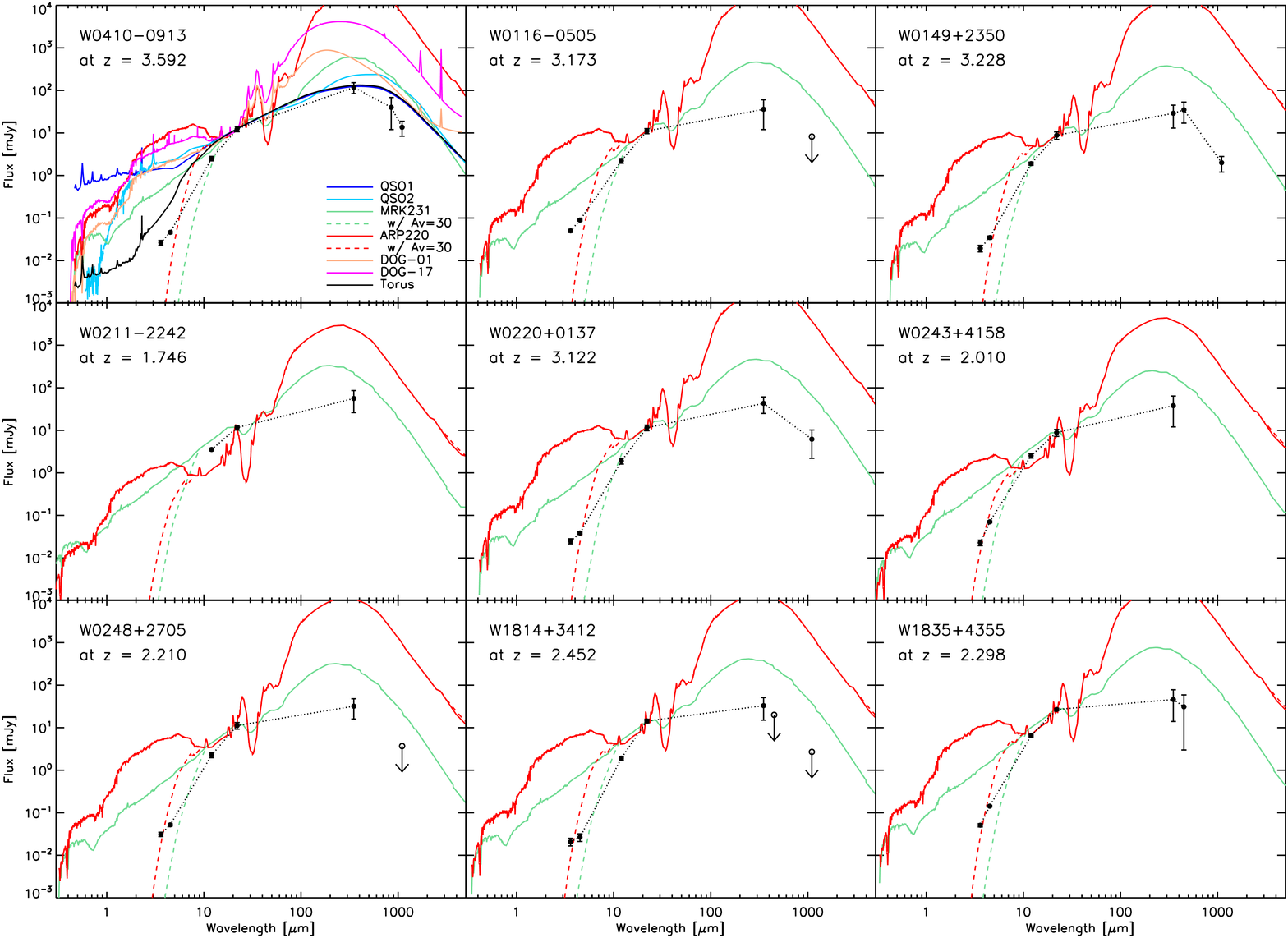}
\caption{
The SEDs for SHARCII detected W12drop galaxies with measured photometry, overlaid on a variety of standard SED templates at their spectroscopic redshifts (Polletta et al. 2006, 2007, Narayanan et al. 2010), normalized at 22 $\mu$m. Additional visual extinction must be added in order to account for the extremely red mid-IR colors of the W12drop galaxies. Black dotted lines in the figure demonstrate the method to connect SED points with power-laws to approximate the total luminosity, as discussed in section 3.  
}
\end{figure}

\begin{figure}
\epsscale{1.0}
\plotone{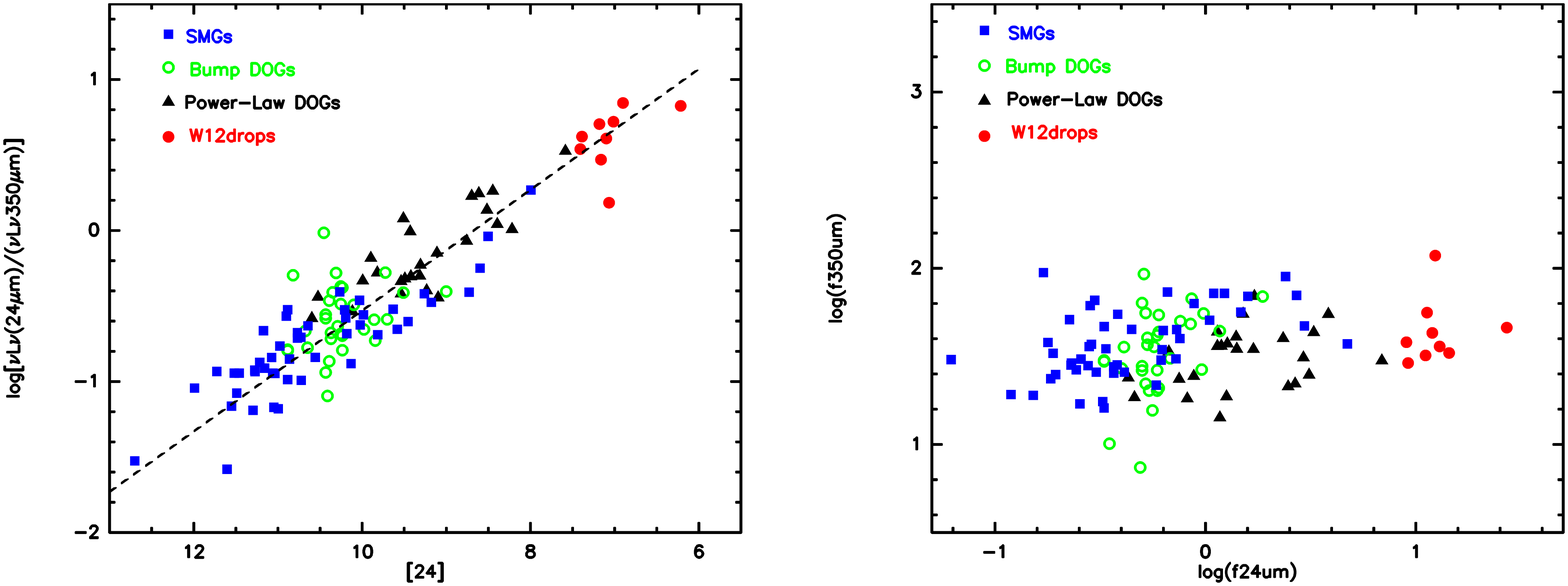}
\caption{{\bf a} (left): The mid-IR to submm luminosity ratios ($\nu$L$\nu$) for SMGs (Magnelli et al. 2012), bump DOGs and power-law DOGs (Melbourne et al. 2012), and W12drops (this work). The dashed line indicates a linear correlation with a fixed slope of unity, and a fitted offset: 
log (F$_{24\mu m}$)=log($\frac{\nu{\rm L}\nu(24\mu m)}{\nu{\rm L}\nu(350\mu m)}$)+0.4. {\bf b} (right): The 350 $\mu m$ emission is similar in all these populations.
}
\end{figure}

\begin{figure}
\epsscale{0.80}
\plotone{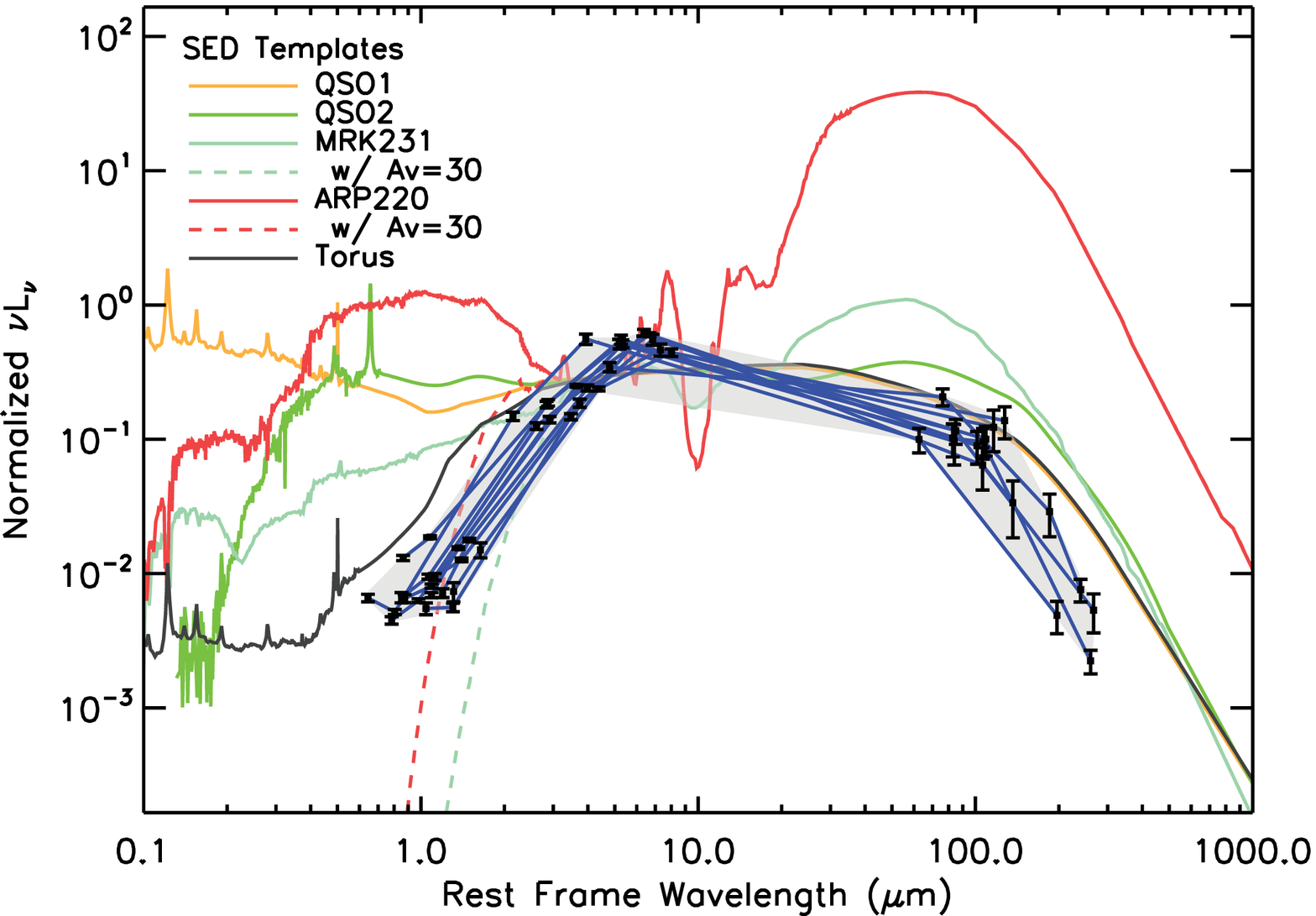}
\caption{$\nu$L$_{\nu}$ units for SEDs in SHARC-II detected W12drop galaxies. The SEDs have been normalized by their total luminosities (derived by connoting SED data points with power-laws), and shifted to the rest frequency frame. All SEDs and galaxy templates are normalized at their flux density at rest frame 5 $\mu$m.
}
\end{figure}

\begin{figure}
\epsscale{0.70}
\rotatebox{270}{\plotone{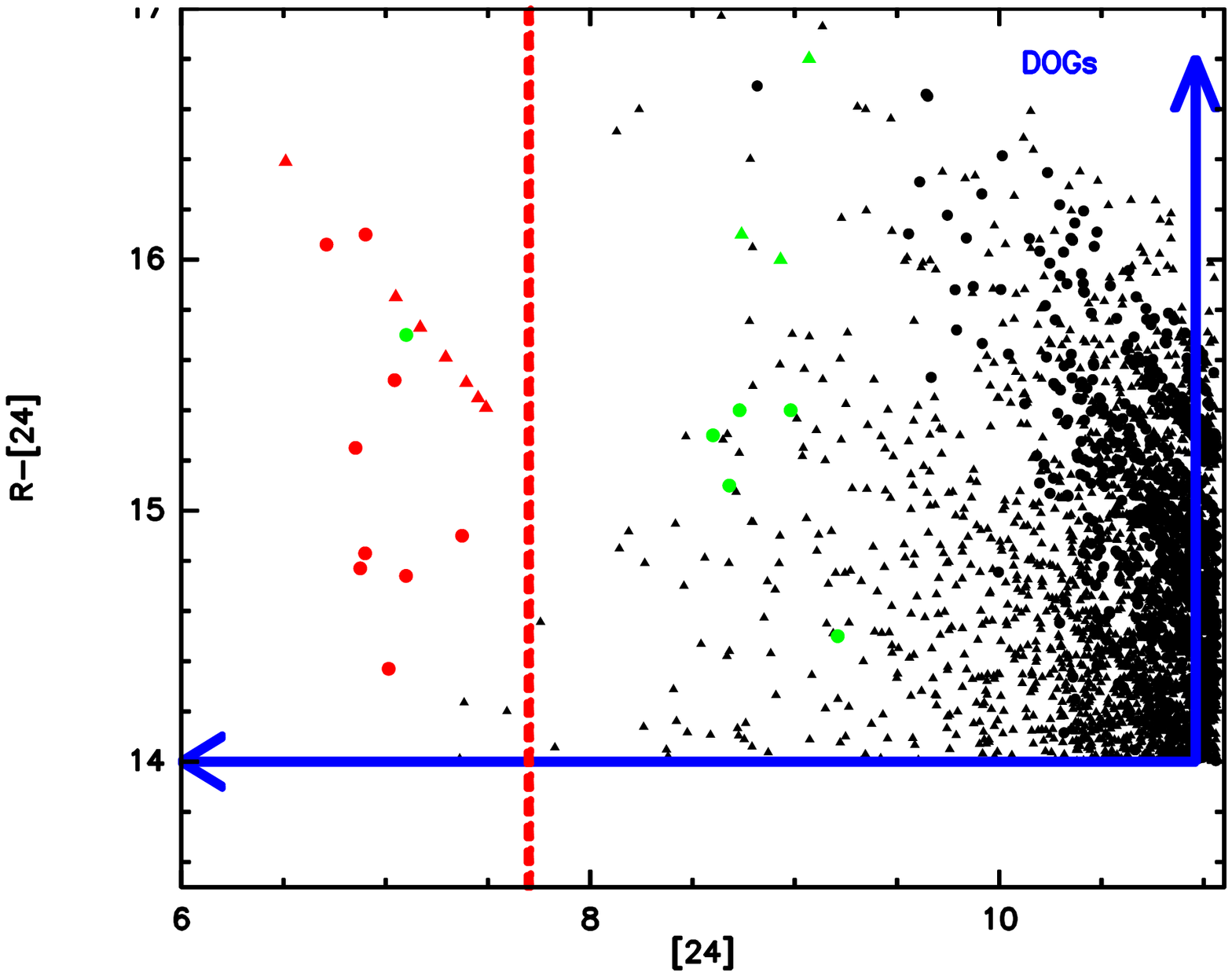}}
\caption{W12drop galaxies in Table 1 (red) compared to DOGs from the Bo$\ddot{\rm o}$tes field (Dey et al. 2008) (black). A subsample of the brightest DOGs which were followed-up by SHARC-II at 350$\mu m$ are indicated in green (Bussmann et al. 2009). Circles represent targets with $R$ band detections and triangles denote targets with $R$ band upper limits . We use SDSS $r$ band and $r$-W4 color to approximate $R$ and $R$-[24] for W12drop galaxies, except for W1814+4512 where we use the $r$-band magnitude reported in Eisenhardt et al. (2012). Blue lines and arrows demonstrate the DOG selection criteria by Dey et al. (2008) , and the red dash line marks the 
lower limit of W4 flux density for the W12drop selection. W12drop galaxies satisfy the DOG classification, but are much brighter at 24 $\mu$m.
}
\end{figure}

\begin{figure}
\epsscale{0.70}
\plotone{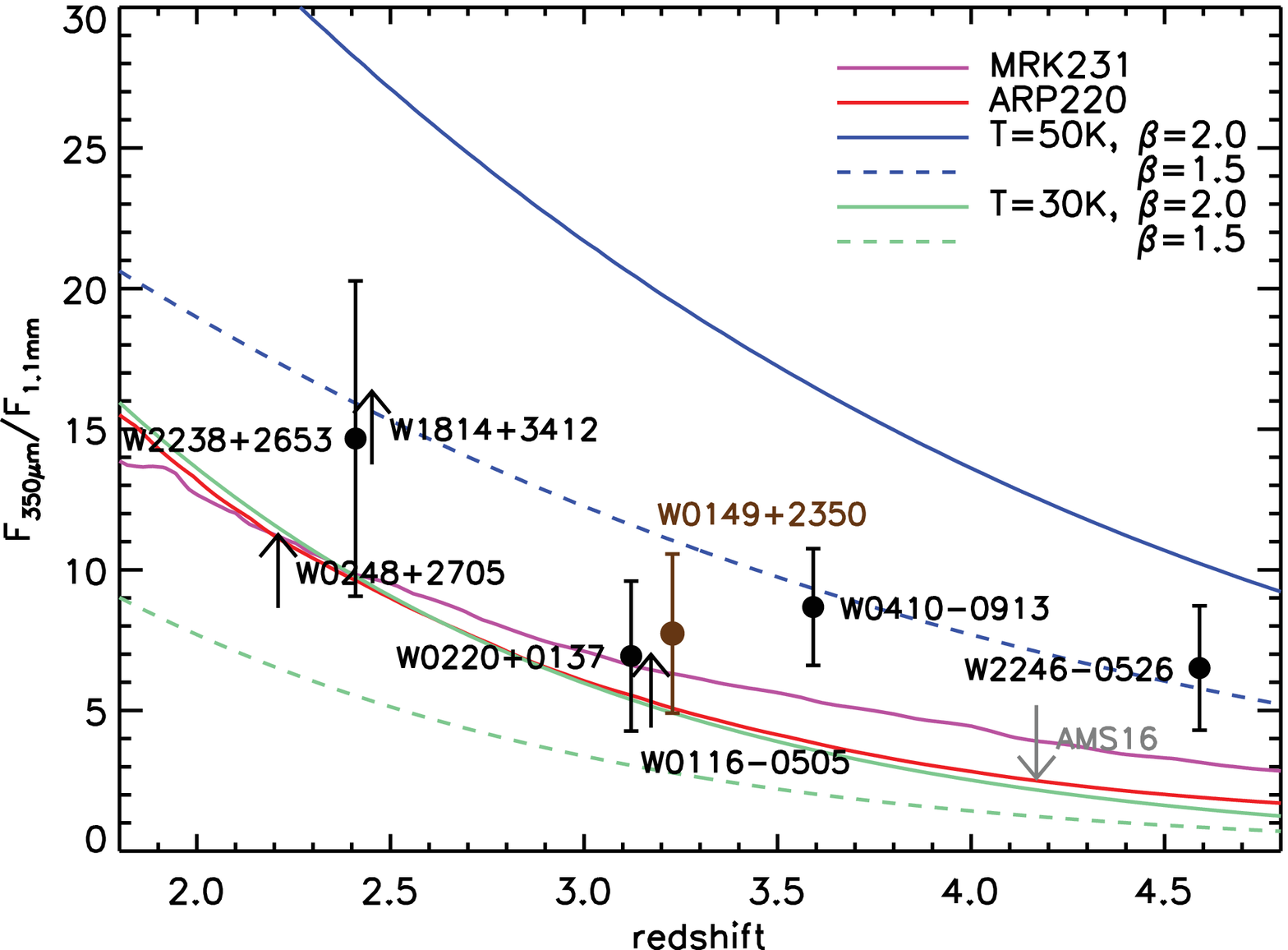}
\caption{Predicted flux ratio of 350$\mu$m to 1.1mm for various dust models as a function of redshift. Upward pointing arrows show values based on 2$\sigma$ limits at 1.1 mm. The 350$\mu$m flux density for W2238+2653 is from Yan et al. (in prep.), and data for W2246$-$0526 is from Tsai et al. (in prep).  For W0149+2350 (brown filled circle), we convert the SMA measurement at 1.3 mm to 1.1 mm flux density, by assuming the emissivity of $\beta$=1.5 and $\beta$=2, then taking the average of the two. For comparison, models of Arp220 and Mrk231 are plotted, and the upper limits of a high-$z$ obscured quasar (AMS16, Mart\'{i}nez-Sansigre et al. 2009) is noted.
}
\end{figure}

\end{document}